# Abnormality Detection inside Blood Vessels with Mobile Nanomachines

Neeraj Varshney, *Student Member, IEEE*, Adarsh Patel, *Student Member, IEEE*,
Yansha Deng, *Member, IEEE*, Werner Haselmayr, *Member, IEEE*, Pramod K. Varshney, *Life Fellow, IEEE*, and
Arumugam Nallanathan, *Fellow, IEEE*

*Abstract*—Motivated by the numerous healthcare applications of molecular communication within Internet of Bio-Nano Things (IoBNT), this work addresses the problem of abnormality detection in a blood vessel using multiple biological embedded computing devices called cooperative biological nanomachines (CNs), and a common receiver called the fusion center (FC). Due to blood flow inside a vessel, each CN and the FC are assumed to be mobile. In this work, each of the CNs perform abnormality detection with certain probabilities of detection and false alarm by counting the number of molecules received from a source, e.g., infected tissue. These CNs subsequently report their local decisions to a FC over a diffusion-advection blood flow channel using different types of molecules in the presence of inter-symbol interference, multi-source interference, and counting errors. Due to limited computational capability at the FC, OR and AND logic based fusion rules are employed to make the final decision after obtaining each local decision based on the optimal likelihood ratio test. For the aforementioned system, probabilities of detection and false alarm at the FC are derived for OR and AND fusion rules. Finally, simulation results are presented to validate the derived analytical results, which provide important insights.

## I. INTRODUCTION

INTERNET of Bio-Nano Things (IoBNT) is gaining significant prominence towards addressing challenging problems in biomedical scenarios [1], where biological cells, produced through synthetic biological processes, are used as biological embedded computing devices or nanomachines to perform sensing, actuation etc. Based on the biological cells and their functionalities in the biochemical domain, biological nanomachines have led to the development of novel applications such as intra-body sensing and actuation, intra-body connectivity control, efficient drug delivery, gene therapy, artificial blood cell production, and human body monitoring by an external health-care provider (see [2]–[4] and the references therein). However, this paradigm poses several research challenges in terms of communication and networking using biochemical infrastructure while enabling an interface to the Internet. Development of efficient and safe techniques for information exchange, interaction, and networking between the biological nanomachines within the IoBNT, is one of the major research challenges. In this context, molecular communication involving transmission and reception of information encoded in molecules, has attracted significant research attention in IoBNT. Molecular communication which is naturally carried out by cells without external influence is ideally suited for above applications especially for abnormality or anomaly detection inside the blood vessels at nano-scale [5].

Recently, some research efforts [6]–[11] have been devoted to addressing abnormality detection such as tumor, cancer, etc. However, none of the works considered the abnormality detection problem in a diffusion-advection blood flow channel, where multiple cooperative biological nanomachines (CNs) and a common receiver or fusion center (FC) also move along with the information molecules with blood flow. This work, therefore, addresses the abnormality detection problem in a blood vessel where each of the mobile CNs are assumed to perform abnormality detection with certain probability of detection and probability of false alarm by counting the number of molecules received from a source, e.g., infected tissue, and report the local decisions to a FC using different types of molecules. Further, simple OR and AND fusion rules are employed at the FC to infer the presence or absence of the abnormality after receiving the local decisions transmitted by each of the CNs over a flow-induced diffusive channel in the presence of inter-symbol interference (ISI), multi-source interference (MSI), and counting errors. These fusion rules are easy to implement due to limited computational capability at the FC. Using the first hitting time model, the probabilities of detection and false alarm at the FC are derived employing OR and AND logic based fusion rules, incorporating the detection performance of the CNs. Here, the first hitting time model best captures the randomness of the arrival time of the molecule due to the Brownian motion of the CNs and FC to make a final decision on abnormality. It is also worth mentioning that in contrast to the passive receiver considered in the existing literature, this work models the receiver nanomachines as fully absorbing receivers [12], [13] which is more practical for health-care applications inside the human body.

## II. SYSTEM MODEL

This work considers cooperative abnormality detection with multiple CNs and a FC inside a blood vessel, i.e., semi-infinite one-dimensional flow-induced fluid medium with constant

N. Varshney, A. Patel, and P. K. Varshney are with the Department of Electrical Engineering and Computer Science, Syracuse University, Syracuse, New York, USA (e-mail:{nvarshne; apatel31; varshney}@syr.edu).

Y. Deng is with the Department of Informatics, King's College London, London, United Kingdom (e-mail: yansha.deng@kcl.ac.uk).

W. Haselmayr is with the Johannes Kepler University Linz, Austria (email: werner.haselmayr@jku.at).

A. Nallanathan is with the Queen Mary University of London, London, United Kingdom (email: nallanathan@ieee.org).

temperature and viscosity, where the length of propagation is large compared to width dimensions. Due to blood flow inside a vessel, all CNs and FC are assumed to be mobile[1] with the flow $v$, i.e., $v_{\text{CN},k} = v_{\text{FC}} = v$, where $v_{\text{CN},k}$ and $v_{\text{FC}}$ denote the velocities of the $k$th CN and the FC, respectively. The diffusion coefficients of the $k$th mobile CN and mobile FC are denoted by $D_{\text{CN},k}$ and $D_{\text{FC}}$, respectively. In this work, the cooperative abnormality detection inside a blood vessel using $K$ CNs is performed as follows.

- **Step 1:** Each CN performs abnormality detection in the $j$th time-slot, $j \in \{1, 2, \cdots\}$, where the $k$th CN obtains the local binary decision based on hypotheses $\mathcal{H}_{0,k}^{\text{CN}}$ and $\mathcal{H}_{1,k}^{\text{CN}}$ with probability of detection ($P_{D,k}^{\text{CN}}[j]$) and probability of false alarm ($P_{F,k}^{\text{CN}}[j]$), where $1 \leq k \leq K$, considering either one of the following scenarios.
  **Scenario 1**: The hypothesis $\mathcal{H}_{0,k}^{\text{CN}}$ defines a normal state at the $k$th CN if the number of molecules $R_k[j]$ received from a source in the $j$th time-slot lies within the threshold range $T_1 \leq R_k[j] \leq T_2$. While the hypothesis $\mathcal{H}_{1,k}^{\text{CN}}$ represents the abnormal state at the $k$th CN such that $R_k[j] < T_1$ or $R_k[j] > T_2$.
  **Scenario 2**: The hypotheses $\mathcal{H}_{0,k}^{\text{CN}}$ and $\mathcal{H}_{1,k}^{\text{CN}}$ define normal and abnormal states if $R_k[j]$ is less or greater than a single threshold $T_1$, respectively.
- **Step 2:** Next, using different types of molecules[2], each CN transmits its local decision obtained in Step 1 to the FC in the subsequent $(j+1)$th time-slot only if it detects an abnormality otherwise it remains silent.
- **Step 3:** After obtaining the local decisions transmitted by all of the CNs over the potentially erroneous diffusive channel using optimal LRT based decision rules, the FC combines the binary decisions and makes a final decision using AND/OR rules to infer the absence or presence of an abnormality inside a blood vessel.

The number of molecules received at the FC corresponding to the transmission of local decision $x_k[j] \in \{0,1\}$ by the $k$th CN in slot $[j\tau, (j+1)\tau]$ can be expressed as

$$\widetilde{R}_k[j+1] = \widetilde{S}_k[j+1] + \widetilde{\mathcal{I}}_k[j+1] + \widetilde{N}_k[j+1] + \widetilde{C}_k[j+1], \quad (1)$$

where $\widetilde{S}_k[j+1]$ represents the number of molecules received in the current $(j+1)$th slot and follows a binomial distribution with parameters $n_k x_k[j]$ and $q_k^0$, i.e., $\mathscr{B}(n_k x_k[j], q_k^0)$, where $n_k$ is the number of type-$k$ molecules transmitted by the $k$th CN for $x_k[j] = 1$ and $q_k^0$ denotes the probability that a transmitted molecule reaches the FC within the current slot. The quantity $\widetilde{N}_k[j+1]$ denotes MSI, i.e., background noise arising due to molecules received from other sources, which can be modeled as a Gaussian distributed random variable with mean $\mu_o$ and variance $\sigma_o^2$ under the assumption that the number of interfering sources is sufficiently large [17]. Also, note that the noise $\widetilde{N}_k[j+1]$ and the number of molecules $\widetilde{S}_k[j+1]$ received from the intended CN are independent [18]. The term $\widetilde{C}_k[j+1]$ denotes the error in counting the type-$k$ molecules at the FC, also termed as the "counting error". This can be modeled as a Gaussian distributed random variable with zero mean and variance that depends on the average number of molecules received as, $\sigma_{c,k}^2[j+1] = \mathbb{E}\{\widetilde{R}_k[j+1]\}$ [18], [19]. The quantity $\widetilde{\mathcal{I}}_k[j+1]$ is the ISI arising in slot $j+1$ due to transmissions in the previous slots and is given as

$$\widetilde{\mathcal{I}}_k[j+1] = \widetilde{I}_k[2] + \widetilde{I}_k[3] + \cdots + \widetilde{I}_k[j], \quad (2)$$

where $\widetilde{I}_k[i] \sim \mathscr{B}(n_k x_k[j-i+1], q_k^{j-i}), 2 \leq i \leq j$ denotes the number of stray molecules received corresponding to the transmission of binary decision $x_k[j-i+1] \in \{0,1\}$ in the $(j-i+2)$th slot. Moreover, the probability $q_k^{j-i}$ that a molecule transmitted by the $k$th CN in slot $i \in \{1, 2, \cdots, j\}$ arrives at FC during time-slot $j$ can be obtained as [20, Eq. (1)]

$$q_k^{j-i} = \int_{(j-i)\tau}^{(j-i+1)\tau} f_k(t; i) dt, \quad (3)$$

where $f_k(t; i)$ is the probability density function (PDF) of the first hitting time, i.e., the time required for a molecule to reach the FC. The PDF $f_k(t; i)$ for a flow-induced diffusive channel considering mobile $k$th CN and FC with flow $v$, i.e., $v_{\text{CN},k} = v_{\text{FC}} = v$, $D_{\text{CN},k} \neq 0$, and $D_{\text{FC}} \neq 0$, is given by[3] [16, Eq. (16)]

$$\begin{aligned}
f_k(t; i) &= \frac{\sqrt{i\tau D_{\text{tot},k} D}}{\pi \sqrt{t} w_k(t;i)} \exp\left(\frac{-(d_{0,k})^2}{4i\tau D_{\text{tot},k}}\right) + \frac{d_{0,k}}{\sqrt{4\pi D(u_k(t;i))^3}} \\
&\quad \times \exp\left(\frac{-(d_{0,k})^2}{4D u_k(t;i)}\right) \operatorname{erf}\left(\frac{d_{0,k}}{2} \sqrt{\frac{tD}{i\tau D_{\text{tot},k} w_k(t;i)}}\right),
\end{aligned} \quad (4)$$

where $u_k(t;i) \triangleq t + i\tau D_{\text{tot},k}/D$ and $w_k(t;i) \triangleq i\tau D_{\text{tot},k} + tD$. The distance $d_{0,k}$ is the Euclidean distance between the $k$th CN and the FC at time $\tau = 0$, $\operatorname{erf}(x)$ denotes the standard error function and the quantities $D_{\text{tot},k}$ and $D$ are defined as, $D_{\text{tot},k} = D_{\text{CN},k} + D_{\text{FC}}$ and $D = D_{\text{FC}} + D_{\text{P}}$ respectively. Further, if the number of molecules released by the $k$th CN satisfy $n_k q_k^0 \geq 5$ and $n_k(1 - q_k^0) > 5$ [18], the binomial distribution for $\widetilde{S}_k[j+1]$ can be approximated by the Gaussian distribution with mean $\mu_k[j+1] = n_k x_k[j] q_k^0$ and variance $\sigma_k^2[j+1] = n_k x_k[j] q_k^0 (1 - q_k^0)$, i.e., $\widetilde{S}_k[j+1] \sim \mathcal{N}(n_k x_k[j] q_k^0, n_k x_k[j] q_k^0 (1 - q_k^0))$ [21]. Similarly, the binomial distribution of $\widetilde{I}_k[i], 2 \leq i \leq j$ can be approximated as

$$\begin{aligned}
\widetilde{I}_k[i] \sim \mathcal{N}(&\mu_{I,k}[i] = n_k x_k[j-i+1] q_k^{i-1}, \\
&\sigma_{I,k}^2[i] = n_k x_k[j-i+1] q_k^{i-1} (1 - q_k^{i-1})).
\end{aligned}$$

Further note that $\widetilde{S}_k[j+1]$ and $\widetilde{I}_k[i], i = 2, 3, \cdots, j$ are independent since the molecules transmitted in different time slots do not interfere with each other [18], [22]. Based on the system model discussed above, the AND and OR logic based rules at the FC are defined as follows.

---

[1]Similar to [14]–[16], the movement of each CN and the FC is modeled as a one dimensional Gaussian random walk. It is assumed that the movement of each CN and the FC does not disrupt the propagation of the information molecules. Moreover, the CNs and the FC can pass each other (see [16] for detailed information).

[2]The molecular propagation from each CN to the FC occurs via Brownian motion with diffusion coefficient $D_{\text{P}}$.

[3]The derived PDF is also verified through particle-based simulations in [16]. It is worth noting that the PDF in (4) is equivalent to the first hitting time PDF [14, Eq.(6)] for diffusion channels without flow and mobile CN and FC. This is due to the fact that the effective flow velocity, i.e., $v - v_{\text{FC}}$, considering the relative motion between the information molecules and the FC, is zero as FC is moving with the same flow $v$ i.e., $v_{\text{FC}} = v$.

- **AND rule:** there is abnormality if the decisions obtained from all the CNs report an abnormal state.
- **OR rule:** there is abnormality if at least one decision obtained at the FC reports an abnormal state.

## III. DETECTION PERFORMANCE ANALYSIS AT FC

Let $\mathcal{H}_0$ and $\mathcal{H}_1$ denote the hypotheses corresponding to the absence and presence of abnormality inside a blood vessel. The average probability of detection $Q_D^l$ and probability of false alarm $Q_F^l$ at the FC corresponding to CN transmissions in time-slots 1 to $l$ are given as

$$Q_D^l = \frac{1}{l} \sum_{j=1}^{l} Q_D[j+1], \quad (5)$$

$$Q_F^l = \frac{1}{l} \sum_{j=1}^{l} Q_F[j+1], \quad (6)$$

where $Q_D[j+1]$ and $Q_F[j+1]$ denote the probabilities of detection and false alarm at the FC corresponding to the transmission by each of the CNs in the $(j+1)$th time-slot. The closed-form expressions for $Q_D[j+1]$ and $Q_F[j+1]$ are derived next for AND and OR fusion rules at the FC.

*1) AND Rule:* The probabilities of detection $Q_D[j+1]$ and false alarm $Q_F[j+1]$ at the FC corresponding to the transmission by each of the CNs in the $(j+1)$th time-slot can be derived as

$$Q_D[j+1] = \Pr(\mathcal{H}_1|\mathcal{H}_1) = \prod_{k=1}^{K} \Pr(\mathcal{H}_{1,k}^{\text{FC}}|\mathcal{H}_1), \quad (7)$$

$$Q_F[j+1] = \Pr(\mathcal{H}_1|\mathcal{H}_0) = \prod_{k=1}^{K} \Pr(\mathcal{H}_{1,k}^{\text{FC}}|\mathcal{H}_0), \quad (8)$$

where $\Pr(\mathcal{H}_{1,k}^{\text{FC}}|\mathcal{H}_1)$ and $\Pr(\mathcal{H}_{1,k}^{\text{FC}}|\mathcal{H}_0)$ can be derived as

$$\Pr(\mathcal{H}_{1,k}^{\text{FC}}|\mathcal{H}_1) = \Pr(\mathcal{H}_{1,k}^{\text{FC}}|\mathcal{H}_{0,k}^{\text{CN}}) \times \Pr(\mathcal{H}_{0,k}^{\text{CN}}|\mathcal{H}_1)$$
$$+ \Pr(\mathcal{H}_{1,k}^{\text{FC}}|\mathcal{H}_{1,k}^{\text{CN}}) \times \Pr(\mathcal{H}_{1,k}^{\text{CN}}|\mathcal{H}_1)$$
$$= P_{F,k}^{\text{FC}}[j+1] \times (1 - P_{D,k}^{\text{CN}}[j])$$
$$+ P_{D,k}^{\text{FC}}[j+1] \times P_{D,k}^{\text{CN}}[j], \quad (9)$$

$$\Pr(\mathcal{H}_{1,k}^{\text{FC}}|\mathcal{H}_0) = \Pr(\mathcal{H}_{1,k}^{\text{FC}}|\mathcal{H}_{0,k}^{\text{CN}}) \times \Pr(\mathcal{H}_{0,k}^{\text{CN}}|\mathcal{H}_0)$$
$$+ \Pr(\mathcal{H}_{1,k}^{\text{FC}}|\mathcal{H}_{1,k}^{\text{CN}}) \times \Pr(\mathcal{H}_{1,k}^{\text{CN}}|\mathcal{H}_0)$$
$$= P_{F,k}^{\text{FC}}[j+1] \times (1 - P_{F,k}^{\text{CN}}[j])$$
$$+ P_{D,k}^{\text{FC}}[j+1] \times P_{F,k}^{\text{CN}}[j], \quad (10)$$

where $P_{D,k}^{\text{FC}}[j+1]$ and $P_{F,k}^{\text{FC}}[j+1]$ denote the probabilities of detection and false alarm at the FC corresponding to the transmission by the $k$th CN in the $(j+1)$th time-slot.

*2) OR Rule:* The probabilities of detection $Q_D[j+1]$ and false alarm $Q_F[j+1]$ at FC can be derived as

$$Q_D[j+1] = \Pr(\mathcal{H}_1|\mathcal{H}_1) = 1 - \prod_{k=1}^{K} \Pr(\mathcal{H}_{0,k}^{\text{FC}}|\mathcal{H}_1), \quad (11)$$

$$Q_F[j+1] = \Pr(\mathcal{H}_1|\mathcal{H}_0) = 1 - \prod_{k=1}^{K} \Pr(\mathcal{H}_{0,k}^{\text{FC}}|\mathcal{H}_0), \quad (12)$$

where $\Pr(\mathcal{H}_{0,k}^{\text{FC}}|\mathcal{H}_1)$ and $\Pr(\mathcal{H}_{0,k}^{\text{FC}}|\mathcal{H}_0)$ are given as

$$\Pr(\mathcal{H}_{0,k}^{\text{FC}}|\mathcal{H}_1) = \Pr(\mathcal{H}_{0,k}^{\text{FC}}|\mathcal{H}_{0,k}^{\text{CN}}) \times \Pr(\mathcal{H}_{0,k}^{\text{CN}}|\mathcal{H}_1)$$
$$+ \Pr(\mathcal{H}_{0,k}^{\text{FC}}|\mathcal{H}_{1,k}^{\text{CN}}) \times \Pr(\mathcal{H}_{1,k}^{\text{CN}}|\mathcal{H}_1)$$
$$= (1 - P_{F,k}^{\text{FC}}[j+1]) \times (1 - P_{D,k}^{\text{CN}}[j])$$
$$+ (1 - P_{D,k}^{\text{FC}}[j+1]) \times P_{D,k}^{\text{CN}}[j], \quad (13)$$

$$\Pr(\mathcal{H}_{0,k}^{\text{FC}}|\mathcal{H}_0) = \Pr(\mathcal{H}_{0,k}^{\text{FC}}|\mathcal{H}_{0,k}^{\text{CN}}) \times \Pr(\mathcal{H}_{0,k}^{\text{CN}}|\mathcal{H}_0)$$
$$+ \Pr(\mathcal{H}_{0,k}^{\text{FC}}|\mathcal{H}_{1,k}^{\text{CN}}) \times \Pr(\mathcal{H}_{1,k}^{\text{CN}}|\mathcal{H}_0)$$
$$= (1 - P_{F,k}^{\text{FC}}[j+1]) \times (1 - P_{F,k}^{\text{CN}}[j])$$
$$+ (1 - P_{D,k}^{\text{FC}}[j+1]) \times P_{F,k}^{\text{CN}}[j]. \quad (14)$$

Now, the closed-form expressions for $P_{F,k}^{\text{FC}}[j+1]$ and $P_{D,k}^{\text{FC}}[j+1]$ can be obtained by formulating the binary hypothesis testing problem using (1) as

$$\mathcal{H}_{0,k}^{\text{FC}} : \widetilde{R}_k[j+1] = \widetilde{\mathcal{I}}_k[j+1] + \widetilde{N}_k[j+1] + \widetilde{C}_k[j+1]$$
$$\mathcal{H}_{1,k}^{\text{FC}} : \widetilde{R}_k[j+1] = \widetilde{S}_k[j+1] + \widetilde{\mathcal{I}}_k[j+1] + \widetilde{N}_k[j+1] \quad (15)$$
$$+ \widetilde{C}_k[j+1].$$

In (15), the number of molecules $\widetilde{R}_k[j+1]$ corresponds to the null and alternative hypotheses following a Gaussian distribution as

$$\mathcal{H}_{0,k}^{\text{FC}} : \widetilde{R}_k[j+1] \sim \mathcal{N}(\widetilde{\mu}_{k,0}[j+1], \widetilde{\sigma}_{k,0}^2[j+1])$$
$$\mathcal{H}_{1,k}^{\text{FC}} : \widetilde{R}_k[j+1] \sim \mathcal{N}(\widetilde{\mu}_{k,1}[j+1], \widetilde{\sigma}_{k,1}^2[j+1]), \quad (16)$$

where the mean $\widetilde{\mu}_{k,0}[j+1]$ and the variance $\widetilde{\sigma}_{k,0}^2[j+1]$ under the null hypothesis $\mathcal{H}_{0,k}^{\text{FC}}$ are calculated as

$$\widetilde{\mu}_{k,0}[j+1] = \sum_{i=2}^{j} \beta_k^{i-1} n_k q_k^{i-1} + \mu_o, \quad (17)$$

$$\widetilde{\sigma}_{k,0}^2[j+1] = \sum_{i=2}^{j} \left\{ \beta_k^{i-1} n_k q_k^{i-1}(1 - q_k^{i-1}) + \beta_k^{i-1}(1 - \beta_k^{i-1}) \right.$$
$$\left. \times (n_k q_k^{i-1})^2 \right\} + \sigma_o^2 + \widetilde{\mu}_{k,0}[j+1], \quad (18)$$

and the probability $\beta_k^{i-1}$ is given as

$$\beta_k^{i-1} = \Pr(x_k[j-i+1] = 1|\mathcal{H}_1)\Pr(\mathcal{H}_1)$$
$$+ \Pr(x_k[j-i+1] = 1|\mathcal{H}_0)\Pr(\mathcal{H}_0)$$
$$= P_{D,k}^{\text{CN}}[i-1]\beta + P_{F,k}^{\text{CN}}[i-1](1 - \beta),$$

where $\beta$ denotes the probability of occurrence of the abnormality. Similarly, mean $\widetilde{\mu}_{k,1}[j+1]$ and variance $\widetilde{\sigma}_{k,1}^2[j+1]$ under the alternative hypothesis $\mathcal{H}_{1,k}^{\text{FC}}$ are derived as

$$\widetilde{\mu}_{k,1}[j+1] = n_k q_k^0 + \widetilde{\mu}_{k,0}[j+1], \quad (19)$$

$$\widetilde{\sigma}_{k,1}^2[j+1] = n_k q_k^0 (2 - q_k^0) + \widetilde{\sigma}_{k,0}^2[j+1]. \quad (20)$$

Employing the above results in the likelihood ratio test (LRT), the optimal test at the FC corresponding to the transmission by the $k$th CN can be seen as [15, Theorem 1]

$$T(\widetilde{R}_k[j+1]) = \widetilde{R}_k[j+1] \underset{\mathcal{H}_{0,k}^{\text{FC}}}{\overset{\mathcal{H}_{1,k}^{\text{FC}}}{\gtrless}} \gamma_k'[j+1], \quad (21)$$

where the optimal decision threshold $\gamma_k'[j+1]$ is given as

$$\gamma_k'[j+1] = \sqrt{\gamma_k[j+1]} - \alpha_k[j+1]. \quad (22)$$

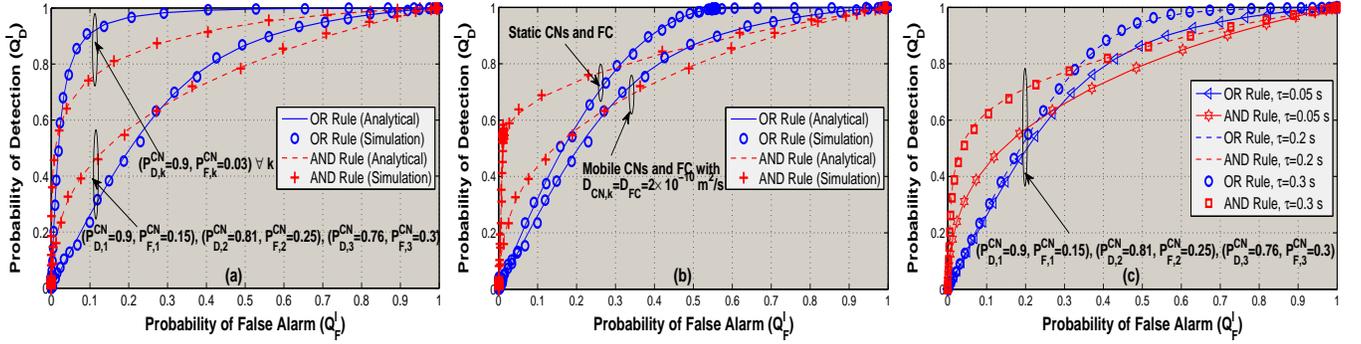

Fig. 1: Detection performance at the FC employing OR and AND fusion rules with (a) different detection performance at the CNs, (b) different mobility conditions with $\tau = 0.05$ s and the detection performance at the CNs as $(P_{D,1}^{CN} = 0.9, P_{F,1}^{CN} = 0.15)$, $(P_{D,2}^{CN} = 0.81, P_{F,2}^{CN} = 0.25)$, $(P_{D,3}^{CN} = 0.76, P_{F,3}^{CN} = 0.30)$, and (c) different values of $\tau$.

In (22), the quantities $\gamma_k[j+1]$ and $\alpha_k[j+1]$ are defined as

$$\alpha_k[j+1] = \frac{\widetilde{\mu}_{k,1}[j+1]\widetilde{\sigma}_{k,0}^2[j+1] - \widetilde{\mu}_{k,0}[j+1]\widetilde{\sigma}_{k,1}^2[j+1]}{\widetilde{\sigma}_{k,1}^2[j+1] - \widetilde{\sigma}_{k,0}^2[j+1]},$$

$$\gamma_k[j+1] = \frac{2\widetilde{\sigma}_{k,1}^2[j+1]\widetilde{\sigma}_{k,0}^2[j+1]}{\widetilde{\sigma}_{k,1}^2[j+1] - \widetilde{\sigma}_{k,0}^2[j+1]} \ln\left[\frac{(1-\beta_k)}{\beta_k}\right]$$
$$\times \sqrt{\frac{\widetilde{\sigma}_{k,1}^2[j+1]}{\widetilde{\sigma}_{k,0}^2[j+1]}} + (\alpha_k[j+1])^2$$
$$+ \frac{\widetilde{\mu}_{k,1}^2[j+1]\widetilde{\sigma}_{k,0}^2[j+1] - \widetilde{\mu}_{k,0}^2[j+1]\widetilde{\sigma}_{k,1}^2[j+1]}{\widetilde{\sigma}_{k,1}^2[j+1] - \widetilde{\sigma}_{k,0}^2[j+1]}.$$

Now, using the above test, the expressions for the $P_{D,k}^{\text{FC}}[j+1]$ and $P_{F,k}^{\text{FC}}[j+1]$ can be obtained as

$$P_{D,k}^{\text{FC}}[j+1] = Q\left(\frac{\gamma'_k[j+1] - \widetilde{\mu}_{k,1}[j+1]}{\widetilde{\sigma}_{k,1}[j+1]}\right), \quad (23)$$

$$P_{F,k}^{\text{FC}}[j+1] = Q\left(\frac{\gamma'_k[j+1] - \widetilde{\mu}_{k,0}[j+1]}{\widetilde{\sigma}_{k,0}[j+1]}\right). \quad (24)$$

## IV. SIMULATION RESULTS AND CONCLUSION

For simulation purposes, the abnormality is considered to occur with probability $\beta = 0.2$ and the various parameters are set as in [22]: the diffusion coefficient $D_P = 242.78 \times 10^{-12}$ m$^2$/s, the number of slots $l = 10$, the number of CNs $K = 3$ with distances $d_{0,1} = 20$ $\mu$m, $d_{0,2} = 15$ $\mu$m, $d_{0,3} = 10$ $\mu$m from the FC at $\tau = 0$, the number of molecules transmitted $n_k = 100$ for $x_k[j] = 1$ $\forall j$, and each CN and FC are assumed to be mobile with diffusion coefficients $D_{\text{CN},k} = D_{\text{FC}} = 2 \times 10^{-10}$ m$^2$/s under flow-induced diffusive channel with drift velocity $v = 3 \times 10^{-3}$ m/s. Moreover, the detection performance at the CNs is considered as mentioned in Fig. 1 and the MSI at the FC is modeled as a Gaussian distributed RV with mean $\mu_o = 10$ and variance $\sigma_o^2 = 10$.

Fig. 1a demonstrates the detection performance at the mobile FC considering different detection performances at the mobile CNs with slot duration $\tau = 0.05$ s. First, it can be observed from Fig. 1a that the analytical values derived in (6) match exactly with the simulation results, thereby validating the derived analytical results. Further, the detection performance at the mobile FC heavily depends on the detection performance of the mobile CNs. The detection performance at the mobile FC significantly improves with the improvement in the detection performance of mobile CNs. One can also observe that at low values of the probability of false alarm ($Q_F^l$), the AND fusion rule outperforms the OR rule. However, as the value of $Q_F^l$ increases, significant increase in performance gain of the OR rule can be observed over the AND rule. For low $Q_F^l$, it is intuitive that the probability of detection ($Q_D^l$) for the AND rule will be better than the OR rule because the AND rule decides $\mathcal{H}_1$ only when all the mobile CNs say $\mathcal{H}_1$. However, for higher values of $Q_F^l$, i.e., each mobile CN is likely to be in error, the increase in the $Q_D^l$ for the AND rule will be more than the OR rule.

Fig. 1b shows the impact of mobility on the detection performance at the FC employing both OR and AND fusion rules, where the diffusion coefficients $D_{\text{CN},k}, D_{\text{FC}}$ are zero for fixed CNs and FC as considered in [1, Fig. 2d]. It can be seen that in comparison to the fixed or static case, the detection performances at the FC under OR and AND rules significantly degrade for the scenario when each CNs and FC are mobile in a flow-induced diffusive medium with $v = 3 \times 10^{-3}$ m/s. This is due to the fact that the probability of a molecule reaching the FC within the current slot, i.e., $q_k^0$ progressively decreases while the ISI from previous slots increases as the diffusion coefficients increase due to mobility. It is also important to note that the cross over point, after which the OR fusion rule performs better than the AND rule, decreases from $(Q_D^l = 0.79, Q_F^l = 0.3)$ to $(Q_D^l = 0.64, Q_F^l = 0.27)$ with the increase in $D_{\text{CN},k}$ and $D_{\text{FC}}$.

Fig. 1c illustrates the detection performance at the FC for different values of slot duration ($\tau$), where each CN and FC are mobile with diffusion coefficient $2 \times 10^{-10}$ m$^2$/s. It is shown that the detection performance at the mobile FC considering OR and AND fusion rules improves as $\tau$ increases from $0.05$ s to $0.2$ ms. However, the detection performance at the mobile FC saturates on further increase in $\tau$. This is due to the fact that the performance at the mobile FC is dominated by the detection performance of the mobile CNs.

### A. Conclusion

This work analyzed the performance of cooperative abnormality detection with multiple CNs and a FC employing OR/ AND fusion rules, where each CN reports its local decision to the FC over a flow-induced diffusive channel that suffers from ISI, MSI and counting errors. Future studies can focus on modeling of the source to CN link. However, it is not straightforward as the first hitting time PDF with a single source and multiple fully absorbing receivers needs to be derived.